\documentclass[unnumsec,webpdf,contemporary,large]{oup-authoring-template}%
\graphicspath{{.}}

\theoremstyle{thmstyleone}%
\theoremstyle{thmstyletwo}%
\theoremstyle{thmstylethree}%

\begin{document}

\journaltitle{PNAS Nexus}
\DOI{DOI added during production}
\copyrightyear{YEAR}
\pubyear{YEAR}
\vol{XX}
\issue{x}
\access{Published: Date added during production}
\appnotes{Paper}

\firstpage{1}

%\subtitle{Subject Section}

\title[]{Surface chiral Abelian topological order on multilayer cluster Mott insulators}

\author[1,$\ast$]{Xu-Ping Yao\ORCID{0000-0002-5431-6049}}
\author[2,$\ast$]{Chao-Kai Li\ORCID{0000-0001-6297-7456}}
\author[3,4,$\dagger$]{Gang v.~Chen\ORCID{0000-0001-9339-6398}}

\address[1]{\orgdiv{Kavli Institute for Theoretical Sciences}, \orgname{University of Chinese Academy of Sciences}, \orgaddress{\postcode{100190}, \state{Beijing}, \country{China}}}
\address[2]{\orgdiv{Key Laboratory of Quantum Materials and Devices of Ministry of Education, School of Physics}, \orgname{Southeast University}, \orgaddress{\postcode{211189}, \state{Nanjing}, \country{China}}}
\address[3]{\orgdiv{International Center for Quantum Materials, School of Physics}, \orgname{Peking University}, \orgaddress{\postcode{100871}, \state{Beijing}, \country{China}}}
\address[4]{\orgname{Collaborative Innovation Center of Quantum Matter}, \orgaddress{\postcode{100871}, \state{Beijing}, \country{China}}}
\address[$\ast$]{These authors contributed equally to this work}
\corresp[$\dagger$]{To whom correspondence should be addressed: \href{email:chenxray@pku.edu.cn}{chenxray@pku.edu.cn}}

\received{Date}{0}{Year}
\revised{Date}{0}{Year}
\accepted{Date}{0}{Year}

%\editor{Associate Editor: Name}

\abstract{The surface states of a symmetry protected topological state can have many possibilities. 
Here we propose a chiral Abelian topological order on a distinct surface 
of a multilayer-stacked cluster Mott insulating system. The first-principle calculation 
and the slave-rotor mean-field theory are applied to study the surface states of the relevant
material system. The angle-resolved photoemission spectroscopic measurement is further suggested  
to detect the anomalous surface fractionalization of the chiral Abelian topological order
on the surface. The connection with real materials is further discussed. 
We expect our results to inspire the interest in the emergent exotic and 
correlation physics among the cluster Mott insulating systems and 
in the interplay between the two different branches of topological phases. } 


% \keywords[Abbreviations]{abbreviation1, abbreviation2, abbreviation3, abbreviation4}

\otherabstract[Significance statement]{The traditional Landau paradigm classifies phases of matter by symmetry and its spontaneous breaking. This picture has been challenged by the discovery of topological phases, which are classified by the global topology of the wavefunction. Two fundamental classes are symmetry-protected topological (SPT) phases and intrinsic topological orders (ITOs), where symmetry plays a distinct role. While their interplay can produce exotic hierarchical states, concrete material platforms remain rare. In this work, we show how a three-dimensional SPT material, Nb$_3$Br$_8$, can host a canonical ITO—a chiral spin liquid—on its surface, driven by specific electron correlations. By revealing this intertwined hierarchy in a layered material, we bridge these two fundamental paradigms, offering testable predictions and explaining recent experiments.}

\maketitle

\section{Introduction}

Topological states of matter 
have revolutionized our understanding of quantum phases, extending classification 
beyond conventional symmetry-breaking paradigms. 
Two fundamental classes emerge, and they are 
symmetry-protected topological (SPT) states~\cite{PhysRevB.80.155131,PhysRevB.87.155114}
and intrinsic topological orders~\cite{PhysRevB.40.7387,PhysRevB.71.045110,PhysRevB.82.155138}. 
SPT phases are distinguished by novel boundary phenomena, such as gapless edge modes 
or symmetry-protected degeneracies, which are robust only in the presence of specific global symmetries.
Their bulk, however, remains gapped and adiabatically connected to a trivial insulator if the symmetry is broken. 
In stark contrast, intrinsic topological order (ITO) represents a more profound departure from classical intuition. 
It is characterized by features that are immune to any local perturbation, 
such as topological ground-state degeneracy dependent on system topology~\cite{PhysRevB.41.9377}, 
fractionalized quasiparticle excitations with anyonic statistics~\cite{doi:10.1142/S0217979290000139}, 
and long-range entanglement~\cite{PhysRevB.82.155138}. 
Crucially, ITO does not rely on symmetry for its protection, 
originating instead from the topological nature of the many-body wavefunction itself.

While conceptually distinct, one being symmetry-enriched and the other fundamentally emergent, 
these two frameworks can intertwine in remarkable ways~\cite{Pesin2010,PhysRevB.109.085137}. 
A particularly fascinating route is realized when a SPT phase hosts an ITO 
on its boundary~\cite{PhysRevX.3.011016,PhysRevB.87.104406,PhysRevX.5.041013,PhysRevX.7.011020}. 
This occurs when the protecting symmetry, essential for the SPT bulk, 
is explicitly broken at the surface. 
The surface, now devoid of the symmetry shield, may generically open a gap. 
However, the non-trivial bulk topology can enforce that this gapped surface state is not 
trivial, and can be a distinct, symmetry-breaking state or, most intriguingly, 
a long-range entangled topological order. In such a scenario, the ITO on the boundary 
is ``seeded" or compelled by the fingerprint of the bulk SPT phase, and 
provides a hierarchical structure and a robust platform to study ITO.

In this Letter, we concretely demonstrate this hierarchical principle in a system 
composed of stacked layers of cluster Mott insulators where the electrons are Mott localized 
in the clusters~\cite{PhysRevB.97.035124,PhysRevResearch.2.043424,PhysRevLett.113.197202,PhysRevB.93.245134,PhysRevB.96.054405,doi:10.1073/pnas.1706769114}. 
The relevant material realization are the cluster magnets 
Nb$_3$Br$_8$~\cite{doi:10.1021/acsnano.9b04392,Date2025} or
Nb$_3$Cl$_8$~\cite{mzyy-thjq,10.1016/j.newton.2025.100292,PhysRevX.13.041049},
and here the electrons of the Nb$_3$ triangular cluster occupy the molecule orbitals and form 
the cluster Mott insulator. This system is isostructural to the Mo$_3$O$_8$-based 
(two-dimensional) cluster magnets~\cite{PhysRevB.97.035124,PhysRevB.93.245134,PhysRevLett.111.217201}
such as LiZn$_2$Mo$_3$O$_8$, Li$_2$InMo$_3$O$_8$, and ScZnMo$_3$O$_8$, 
and thus share many similar physics~\cite{Sheckelton2012,PhysRevB.93.245134,PhysRevB.89.064407,PhysRevB.96.054405,PhysRevResearch.2.043424,PhysRevLett.120.227201,GALL201399}. 
As far as the universal aspect of the physics is concerned, 
one defining characteristic of cluster Mott insulators 
is the formation of specific intra-layer electron clusters.

More crucially for Nb$_3$Br$_8$ or Nb$_3$Cl$_8$~\cite{doi:10.1021/acsnano.9b04392,Date2025},
the inter-layer tunnelling is structured such that the physics 
along the stacking direction 
is effectively described, at low energies, 
by the Su-Schrieffer-Heeger (SSH) model~\cite{RevModPhys.60.781}. 
As a fundamental paradigm of an SPT phase, the SSH chain is protected by a chiral 
%(or sublattice) 
symmetry. In the three-dimensional stacked architecture, 
this SSH physics is then extended, elevating the system into 
a three-dimensional (weak) SPT phase~\cite{PhysRevB.87.155114}. 
Its non-trivial bulk directly dictates the existence of 
protected gapless two-dimensional surface states. 
The ultimate fate of these surface states, however, is critically dependent 
on the specific surface termination and the role of local interactions.

The key realization is that for a particular surface termination,    
the effective theory for these SPT-mandated anomalous surface states maps onto 
a single-band Hubbard model on a triangular lattice and behaves as if it is a monolayer system. 
Recent theoretical and numerical progresses strongly 
suggest that the triangular lattice Hubbard model, at the intermediate correlation, 
can host a chiral spin liquid (CSL) ground state~\cite{PhysRevX.10.021042,PhysRevLett.127.087201}, 
a canonical example of an intrinsic topological order. 
This state is characterized by time-reversal symmetry breaking chiral edge mode, 
fractionalized anyonic excitations, and a non-zero spinon Chern number. 
More remarkably, a superconducting diode effect was observed in the
superconducting Josephson junction NbSe$_2$/Nb$_3$Br$_8$/NbSe$_2$ where Nb$_3$Br$_8$
serves as a Mott insulating barrier~\cite{Wu2022}, and
that seems compatible with CSL. In our specific construction, 
we propose the ``fingerprint" of the bulk SPT phase, 
mediated through the correlation of the surface Hubbard model, 
catalyzes the emergence of stable chiral Abelian topological order. 
The purpose of this Letter is to identify the possible existence of 
this surface topological order theoretically and propose the relevant 
surface detection scheme.

\begin{figure*}[b]
    \centering
	\includegraphics[scale=2.0]{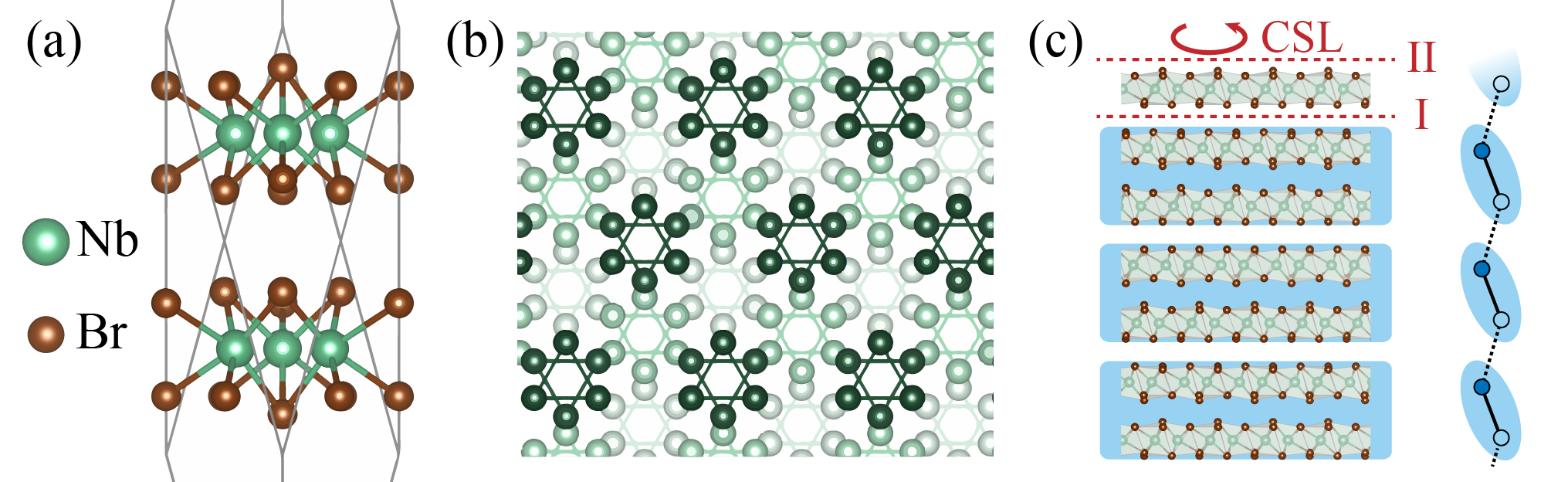}
	\caption{The crystal structure of Nb$_3$Br$_8$. 
	(a) The primitive cell consists of a bilayer unit. 
	(b) Within the bilayer unit, the centers of the triangles from the two layers are aligned along the c-axis, 
	forming bilayer clusters. Different shades refer to bilayer clusters at different depths. 
	(c) The side view of the stacking pattern (left) is reminiscent of the SSH model (right). 
	The two different types of surface termination are labeled by I and II. A CSL can exist on the type-II surface.
    Crystal structures are drawn by VESTA software~\cite{Momma:db5098}.
	}%
	\label{fig: crystal_structure}
\end{figure*}

\section{Results}

\subsection{Electronic Structure of Nb$_3$Br$_8$}

Nb$_3$Br$_8$ is a layered van der Waals (vdW) material with a rhombohedral primitive cell 
[Fig.~\ref{fig: crystal_structure}(a)]~\cite{Simon1966,Habermehl2010}. 
In each layer, the magnetic Nb ions form a breathing-kagom\'e lattice. 
The kagom\'e lattice consists of two sets of triangles, 
and in the breathing kagom\'e lattice, the bond lengths 
in one set of the triangles are shorter than those in the other set
[see Fig.~\ref{fig: crystal_structure}(a)]. The breathing parameter of Nb$_3$Br$_{13}$ 
is 1.46 and is much larger than LiZn$_2$Mo$_3$O$_8$ and Li$_2$InMo$_3$O$_8$. 
The smaller triangles then contribute the building clusters for the localized electrons 
in Nb$_3$Br$_{13}$, and these clusters themselves arrange in a triangular lattice pattern. 
In the crystal field of NbBr$_6$ octahedra, the low-lying $t_{2g}$-orbitals 
in the three Nb atoms in a cluster form the molecular orbitals. 
The highest occupied one is a $2A_1$ molecular orbital filled by one electron, 
similar to the situation of LiZn$_2$Mo$_3$O$_8$ and 
Nb$_3$Cl$_8$~\cite{PhysRevX.13.041049,Sheckelton2017,Haraguchi2017}. 
Owing to the strong breathing structure and the spin degeneracy of 
the $2A_1$ molecular orbital, a monolayer of Nb$_3$Br$_8$ can be effectively 
viewed as a half-filled triangular lattice with a single molecular band. 

The low-temperature (${ T < 382~\text{K} }$) phase of Nb$_3$Br$_8$
consists of bilayer units, in which the two constituent layers are related by an inversion. 
Within the bilayer, the centers of the triangular clusters in the upper layer lie right  
on top of the centers of the triangular clusters in the lower layer,  
forming a bilayer cluster. Because the molecular orbital is localized 
at the center of each cluster, the bilayer can be regarded as 
an AA-stacked triangular lattice. The bilayer units are further ABC-stacked 
along the c-direction (see Fig.~\ref{fig: crystal_structure}(b)), 
resulting in a space group $R\bar{3}m$. One expects that the hopping matrix elements 
between the bilayers to be smaller than those within a bilayer. 
Such comparison of the hopping strengths is reminiscent of the SSH model 
along the stacking direction (see Fig.~\ref{fig: crystal_structure}(c)).

The ground state of a one-dimensional SSH model is an insulator. 
In the three-dimensional Nb$_3$Br$_8$, there are additional in-plane hoppings. 
Because the Nb ions in each layer form clusters, however,
the distance between clusters is large, resulting in an small in-plane hopping 
as compared with the out-of-plane hopping between the AA-stacked molecular orbitals. 
The effect of the alternating interlayer hoppings still 
dominates the relatively weak in-plane dispersion. 
Hence, the three-dimensional bulk band structure will still be insulating.

Because of the difference of stacking fashions within a bilayer unit and between them, 
there are two types of surfaces. When the surface terminates at a full bilayer unit, 
it is called a type-I surface. In contrast, if the surface terminates in the middle of a bilayer unit, 
it is called a type-II surface (Fig.~\ref{fig: crystal_structure}(c)). 
In view of a one-dimensional SSH model, these two types of termination are 
topologically distinct in their band structure, which can be exhibited by 
the existence or absence of the zero modes at the real-space boundaries.
In three-dimensional Nb$_3$Br$_8$, the similar physics can be observed 
at the two types of surfaces. 

\begin{figure}[b]
    \centering
	\includegraphics[width=1.0\linewidth]{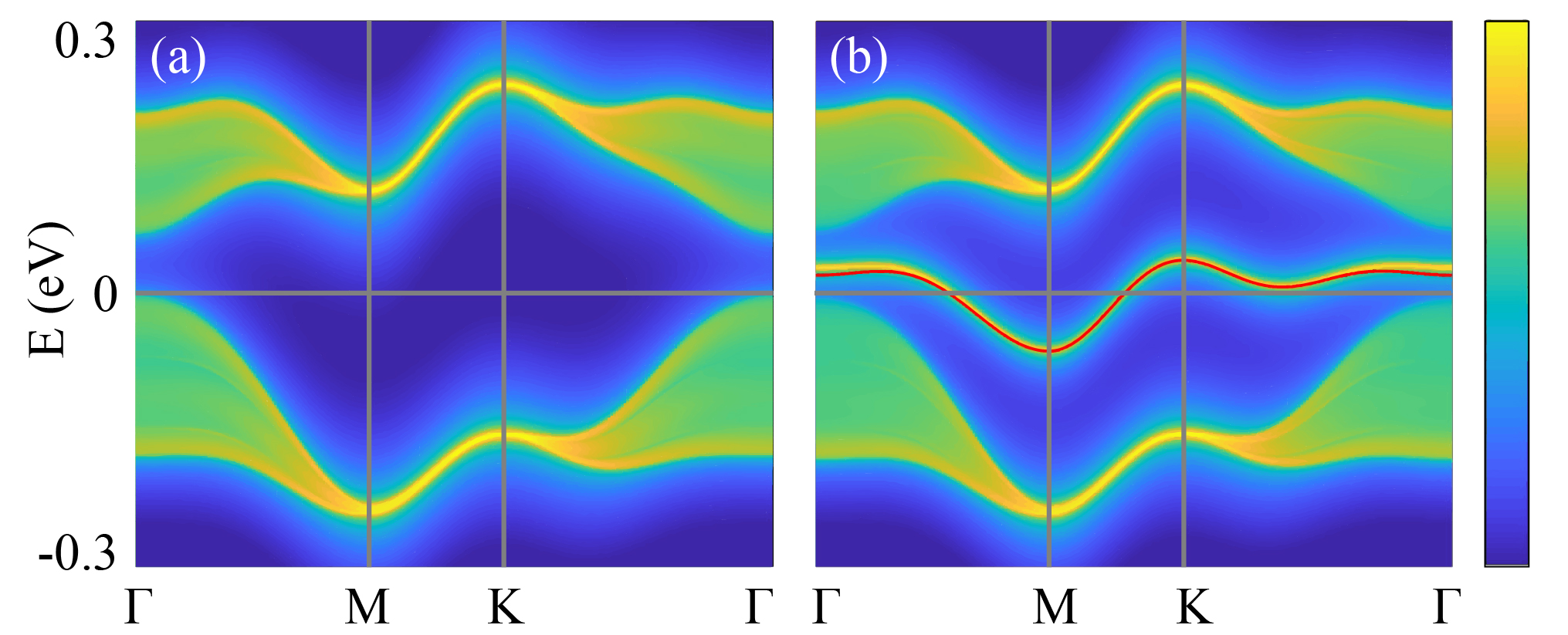}
	\caption{(a) The spectral function of the type-I surface. 
	(b) The spectral function of the type-II surface. 
	The red curve is the band dispersion of a monolayer tight binding model. 
	It agrees well with the surface metallic band.
    The surface spectral functions
	are obtained from the surface Green's functions calculated by the recursive 
	algorithm described in Ref.~\cite{Sancho1985}.}%
	\label{fig: band_structures_surface}
\end{figure}

This observation is corroborated by our first-principles calculations. 
The band structure shows that Nb$_3$Br$_8$ is a band insulator in the bulk, 
with a direct band gap of about $ 70~\text{meV} $ located at the Brillouin zone boundary 
(see Fig.~S1 in \textit{SI Appendix}).
The surface spectral functions of electrons are calculated 
by implementing the recursive Green's function method~\cite{Sancho1985}, 
and the results for the two types of surfaces are shown 
in Figs.~\ref{fig: band_structures_surface}(a-b), respectively. 
It can be clearly seen that the type-I surface remains gapped,  
similar to the bulk band structure. For the type-II surface, 
however, there is a weakly dispersive band crossing the Fermi level, 
suggesting a metallic type-II surface. Such metallic surface state 
corresponds to the zero mode in the SSH model. 
Moreover, we find that the metallic surface state of the type-II surface 
can be well captured and reproduced by the tight-binding model of a Nb$_3$Br$_8$
monolayer (see Fig.~\ref{fig: band_structures_surface}(b)). Therefore, 
in the following we use a tight-binding model of monolayer Nb$_3$Br$_8$ 
as the non-interacting part of the model for the surface state.

The monolayer tight-binding model describing a triangular lattice 
with hoppings up to the third nearest-neighbors is written as follows,
\begin{equation}
	H_t=-t_{1}\sum_{\langle ij \rangle}c_{i\sigma}^{\dagger}c_{j\sigma}
	-t_{2}\sum_{\langle\langle ij \rangle\rangle}c_{i\sigma}^{\dagger}c_{j\sigma}-t_{3}\sum_{\langle\langle\langle ij \rangle\rangle\rangle}c_{i\sigma}^{\dagger}c_{j\sigma},
	\label{eq1}
\end{equation}
where $c_{i\sigma}^{\dagger}$ ($c_{i\sigma}$) creates (annihilates) 
an electron with spin $\sigma$ at site $i$. 
The Wannier function was constructed from the Kohn-Sham wave 
functions of the band structure calculation, and the hopping parameters 
were found to be $t_1 = -5.0~\text{meV}$, $t_2 = -5.5~\text{meV}$, 
and $t_3 = 6.8~\text{meV}$. 
The corresponding band structure is shown by the red curve 
in Fig.~\ref{fig: band_structures_surface}(b), as mentioned above.

We note that these hopping parameters are heavily suppressed due to the large spatial separation between neighboring molecular orbitals. 
In contrast, the Hubbard $U$ interaction, calculated via the constrained random phase approximation, is on the order of eV, thereby rendering Nb$_3$Br$_8$ a strongly correlated insulator~\cite{wr7w-nfhg}
This strong correlation scenario is further supported by the angle-resolved photoemission spectroscopy (ARPES) measurements~\cite{Date2025,wr7w-nfhg}, which resolve both the lower and upper Hubbard bands, providing direct evidence of prominent many-body effect.
In Ref.~\cite{PhysRevB.107.035126}, the authors used the slave boson theory to 
calculate the metal-insulator transition of monolayer Nb$_3$X$_8$ (X=Cl, Br, and I), and found 
that a $ U $ interaction in $ 10^{2} $~meV range is enough to drive a 
metal-insulator transition. Thus, it is expected that the metallic surface state 
on the type-II surface will be driven into a Mott insulating state after 
the Coulomb interaction is properly taken into account.

\subsection*{Slave rotor mean-field theory}

To capture the interaction physics on the monolayer of type-II surface in Nb$_3$Br$_8$, 
we begin with the half-filled Hubbard Hamiltonian by introducing 
correlation to the tight-binding model in~\eqref{eq1}, 
\begin{equation}\label{eq:hubbard}
    H = - \sum_{ij,\sigma} (t_{ij} c_{i\sigma}^{\dagger}c_{j\sigma} + \text{h.c.}) 
        + \frac{U}{2} \sum_i (n_{i} - 1)^2,   
\end{equation}
where $n_{i} = \sum_{\sigma}c_{i\sigma}^{\dagger}c_{i\sigma}$ 
is the electron number operator, and $U$ denotes the on-site Coulomb repulsion. 
Up to the 3rd-nearest-neighbor electron hopping, large-scale density-matrix renormalization group calculations reveal that the triangular-lattice Hubbard model could host a gapless chiral spin liquid in the large-$U$ limit~\cite{PhysRevB.100.241111}. 
As $U$ decreases, enhanced quantum fluctuations are anticipated to stabilize non-magnetic ground states over long-range-orders.
To treat the correlation in the weak-to-intermediate $U$ regime
and capture the non-magnetic ground state suggested by experiments~\cite{doi:10.1021/acsnano.9b04392,Wu2022}, 
we employ the slave-rotor approximation~\cite{PhysRevB.70.035114,PhysRevLett.95.036403}. In this framework, the electron operators is decomposed 
into a fermionic spinon operator $f^{\dagger}$ carrying the spin degree of freedom 
and a bosonic rotor $\Phi$ representing the charge
$c_{i\sigma}^{\dagger} = f_{i\sigma}^{\dagger} \Phi_i, c_{i\sigma}=f_{i\sigma} \Phi_i^{\ast}$, 
where $\Phi_i=e^{\imath\theta_i}$. The charge quantum number corresponds to an angular momentum 
$L_i = -\imath \partial_{\theta_i}$, satisfying $[\theta_i, L_j]=\imath\delta_{ij}$, 
and is constrained by $L_i=\sum_{\sigma}f_{i\sigma}^{\dagger}f_{i\sigma}-1$. 
The Hubbard Hamiltonian is reformulated as 
\begin{multline}
    H = \sum_{ij,\sigma} t_{ij} e^{\imath(\theta_i-\theta_j)} f_{i\sigma}^{\dagger} f_{j\sigma} 
    + \text{h.c} - \sum_{i\sigma} (h_i + \mu_i) f_{i\sigma}^{\dagger}f_{i\sigma} \\
    + \sum_i \frac{U}{2}L_i^2 + h_i L_i + h_i + \mu_i,
\end{multline}
where the Lagrange multipliers $h_i$ and $\mu_i$ have been introduced to 
recover the physical Hilbert space. The non-quadratic terms between spinons 
and rotors can be further decoupled via the Hubbard-Stratonovich transformation. 
The decomposition introduces two auxiliary fields ${\Delta_{ij} = -t_{ij} \langle\Phi_i\Phi_j^{\ast}\rangle}$ 
and $\chi_{ij} = -t_{ij} \sum_{\sigma} \langle f_{i\sigma}^{\ast}f_{j\sigma} \rangle$ 
and yields the mean-field Hamiltonian for the spinons and the charge rotors
\begin{align}
    H_s & = \sum_{ij,\sigma} \Delta_{ij} f_{i\sigma}^{\dagger}f_{j\sigma} 
    + \text{h.c.} + \sum_i \mu_i (1 - f_{i\sigma}^{\dagger}f_{i\sigma}), 
    \label{eq:MF_spinon} \\
    H_r & = \sum_{i\sigma} \chi_{ij}^{\ast} \Phi_i^{\ast} \Phi_j 
    + \text{h.c.} + \sum_i \frac{U}{2} L_i^2, 
\end{align}
supplemented by a constant $\sum_{ij} \Delta_{ij}\chi_{ij}/t_{ij}$. 
The auxiliary fields are complex in general whose phases are associated the U(1) gauge field. 
In momentum space, the spinon mean-field Hamiltonian $H_s$ is diagonalized by a unitary transformation
\begin{equation}
    \mathcal{U}^{\dagger}_{\boldsymbol{k}} H_s(\boldsymbol{k}) \mathcal{U}_{\boldsymbol{k}} = \mathrm{diag} \{ \xi_{\boldsymbol{k},1}, \xi_{\boldsymbol{k},2}, \ldots, \xi_{\boldsymbol{k},n} \}, 
\end{equation} 
where $n$ is the sublattice index. 
To search the saddle-point solutions that strictly satisfy the local constraints 
${\sum_{\sigma} f_{i\sigma}^{\dagger}f_{i\sigma} = 1}$ for the spinons, 
we employ the self-consistent minimization algorithm developed in 
Refs.~\cite{PhysRevLett.103.135301,PhysRevB.84.174441} and 
its extensions in Refs.~\cite{PhysRevA.93.061601,PhysRevB.105.024401}. 
Once a solution satisfying the spinon local constraints is found, it gives an auxiliary field $\chi_{ij}$ via the spinon correlator $\langle f_{i\sigma}^{\dagger}f_{j\sigma} \rangle$, which in turn specifies the rotor Hamiltonian $H_r$.  
For the rotor Hamiltonian $H_r$, the unimodular condition ${\Phi_i^{\ast}\Phi_i = 1}$ 
gives an implicit local constraint, enforced by the other Lagrange multiplier $\lambda_i$. 
We adopt a uniform saddle-point approximation in the rotor sector by setting ${\lambda_i = \lambda}$. 
Under this approximation, the rotor correlator is evaluated as
\begin{equation}
    \langle \Phi_{\boldsymbol{k}\alpha}^{*}\Phi_{\boldsymbol{k}\beta} \rangle = \sum_{n} \mathcal{V}_{\boldsymbol{k}\alpha n}^{*}\mathcal{V}_{\boldsymbol{k}\beta n} \frac{U}{\sqrt{2U(\epsilon_{\boldsymbol{k} n} + \lambda)}},
\end{equation}
where $n$, $\alpha$, and $\beta$ index the sublattice bands. 
$\epsilon_{\boldsymbol{k}}$ and $\mathcal{V}_{\boldsymbol{k}}$ are eigenvalues and transformation matrices that diagonalize the the quadratic part of the rotor action (see detailed derivations in \textit{SI Appendix}).
The self-consistent equation enforcing the relaxed global constraint over all $N_s$ sites is then 
\begin{equation}
    \frac{1}{N_s} \int_{\boldsymbol{k}\in\text{BZ}} d^2\boldsymbol{k} \sum_{n} \langle \Phi_{\boldsymbol{k} n}^{*}\Phi_{\boldsymbol{k} n} \rangle = 1.
\end{equation}

\subsection*{Saddle-point solutions for monolayer Nb$_3$Br$_8$}

\begin{figure*}[t]
    \centering
    \includegraphics[]{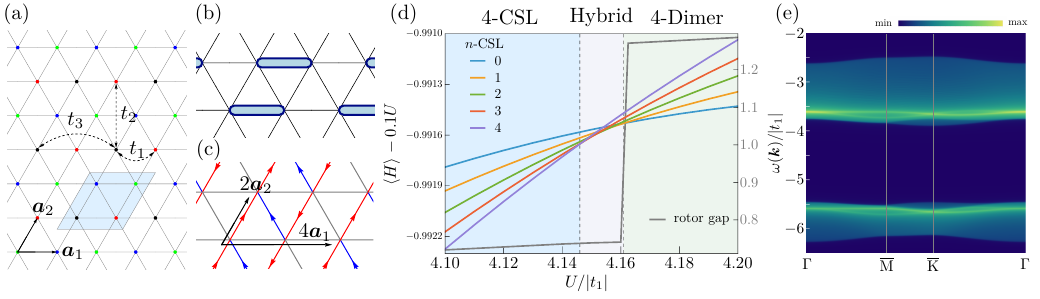}
    \caption{
    (a) Monolayer electron hoppings and four sublattices (shaded diamond) defined by $t_3$.
    (b) One of sublattice dimer states. 
    (c) Sublattice CSL realizing the $\pm\pi/2$ flux pattern on each triangular plaquette. 
    Grey, red, and blue arrows represent $0$, $\pi/2$, and $\pi$ phases on $\Delta_{ij}$, respectively.
    (d) Slave-rotor mean-field phase diagram obtained on an enlarged unit cell up to a $6 \boldsymbol{a}_1 \times 6 \boldsymbol{a}_2$ geometry with periodic boundary conditions. A linear-$U$ term has been added to energy for clarity. 
    Gray line represents the ground-state rotor gap. 
    (e) Electron spectral function $A(\omega \leq 0, \boldsymbol{k})$ for $U/|t_1|=4.0$ at zero-temperature limit.
    The high-symmetry points are defined in the reduced surface Brillouin zone. 
    }%
    \label{fig:monolayer}
\end{figure*}

With the hopping parameters for the monolayer Nb$_3$Br$_8$ and the energy unit $|t_1|$, 
the slave-rotor mean-field phase diagram is presented in Fig.~\ref{fig:monolayer}(d). 
A Fermi liquid phase, originating from the non-interacting limit, 
persists up to the Mott transition at ${U_c/|t_1| \approx 2.24}$, 
as determined by tracing the gap of the charge rotor. 
Beyond the Mott transition, the self-consistent minimization algorithm 
reveals a low-energy subspace characterized by vanishing auxiliary fields 
$\Delta_{ij}$ and $\chi_{ij}$ on the first and second nearest-neighbor bonds,
leaving only the strongest ones on the third nearest-neighbor bonds. 
Similar behaviour has been reported in the square-lattice Hubbard model 
within the same framework~\cite{PhysRevB.83.134515}. 
As illustrated in Fig.~\ref{fig:monolayer}(a), these third nearest-neighbor 
bonds partition the original triangular lattice into four independent sublattices 
(diamond-shaped shaded region). 
At the mean-field level, these four sublattices are effectively decoupled,
each forms a $2\boldsymbol{a}_1 \times 2\boldsymbol{a}_2$ enlarged triangular lattice 
defined by $\Delta_{ij} \propto t_3$ and $\chi_{ij} \propto t_3$. 
On the triangular sublattice, there are two distinct states, i.e.
a dimer state or a U(1) CSL of the Kalmeyer-Laughlin type. 
The dimer state is massively degenerate, 
comprising a manifold spanned by all possible perfect dimer covering. 
An ordered dimer configuration, as presented in Fig.~\ref{fig:monolayer}(b), 
is expected to remain stable beyond the mean-field approximation. 
The U(1) CSL is a Chern insulator of spinons and is realized 
by the mean-field ansatz $\Delta_{ij} = |\Delta|e^{\imath a_{ij}}$ 
with $|\Delta|$ spatially uniform. 
The bond phase $a_{ij}$ plays the role of a fluctuating U(1) gauge field. 
It generates a ${\pm \pi / 2}$ flux on every triangular plaquette for the sublattice
as shown in Fig.~\ref{fig:monolayer}(c). 

While the spinon spectrum is fully gapped in both states, 
the occupied spinon bands in the CSL for each sublattice carry Chern number $\pm 1$, 
whereas those in dimer state are topologically trivial.
Specifically, the coupling between the gapped spinons and the emergent U(1) gauge field $a_{\mu}=(a_0,a_x,a_y)$ is governed by the Euclidean partition function
\begin{equation}
    \mathcal{Z}_s = \int \mathcal{D}[a, f^{\dagger}, f] e^{-\int d\tau \left\{\sum_{i\sigma}f_{i\sigma}^{\dagger}(\partial_{\tau} - \imath a_0)f_{i\sigma} + H_s[f^{\dagger},f,a]\right\}}.
\end{equation}
By integrating out the gapped spinon fields, one can obtain the effective action for the gauge field
\begin{align}
    \mathcal{S}_{s,\text{eff}}[a] = - \mathrm{Tr} \ln [\mathcal{G}_{s}^{-1} - a_{\mu}j^{\mu}],
\end{align}
where $\mathcal{G}_{s}$ is the bare spinon Green's function and $j^{\mu}$ is the spinon current operator.
Treating the gauge fluctuations as long-wavelength collective excitations, we can expand the effective action up to the second order in $a_{\mu}$ 
\begin{equation}
    \mathcal{S}_{s,\text{eff}}[a] = \mathcal{S}_{s,\text{eff}}^{0} + \frac{1}{2} \mathrm{Tr}(\mathcal{G}_s a_{\mu} j^{\mu} \mathcal{G}_{s} a_{\nu} j^{\nu})  + \mathcal{O}(a^3).
\end{equation}
Transforming to the momentum space $q = (\imath\omega_{n},\boldsymbol{q})$, the second-order term reads
\begin{equation}
    \mathcal{S}_{s,\text{eff}}^{(2)}[a] = \frac{1}{2} \int\frac{d^3 q}{(2\pi)^3} a_{\mu}(-q) \Pi^{\mu\nu} (q) a_{\nu}(q),
\end{equation}
where $\Pi^{\mu\nu}(q)$ is the spinon vacuum polarization tensor. 
In the static and long-wavelength limit, the system breaks time-reversal symmetry, and the leading-order antisymmetric part of the polarization tensor is strictly proportional to the Chern number of the occupied spinon bands~\cite{PhysRevB.39.11413}
\begin{equation}
    \Pi^{\mu\nu} (\boldsymbol{q}) = - \imath \frac{C}{2\pi} \epsilon^{\mu\nu\lambda} q_{\lambda} + \mathcal{O}(q^2).
\end{equation}
Substituting this back and performing the Fourier transformation to real space, it explicitly explicitly gives the topological Chern-Simons term in the Euclidean action 
\begin{equation}
    \mathcal{S}_{\text{CS}}[a] = \imath \frac{C}{4\pi} \int d\tau d^2\boldsymbol{r} \epsilon^{\mu\nu\lambda} a_{\mu} \partial_{\nu} a_{\lambda}. 
\end{equation}
This action describes the dynamics of a $2+1$ dimensional U(1) gauge field in the continuum limit and determines the topological properties of the CSL state, including the long-range entanglement. 
In fact, it is the Chern-Simons term that attaches a flux to the spinon, transforming it into an anyon with fractional statistics. 
In the rotor sector, the auxiliary field $\chi_{ij}$ follows the same spatial patterns according to the spinon dimer or CSL, 
although the induced flux carries the opposite sign in the CSL. 

The complete decoupling of the four sublattices allows each to 
independently adopt either a CSL or a dimer state in the spinon sector, 
leading to five distinct phases on the original triangular lattice. 
Each phase can be characterized by the number $n$ (ranging from 0 to 4) 
of sublattices in a CSL, as labelled by $n$-CSL in Fig.~\ref{fig:monolayer}(d).
Figure~\ref{fig:monolayer}(d) also displays the system energies and the ground-state rotor gap. 
For strong interactions ${U/|t_1| \gtrsim 4.161}$, 
the dimer state is favored on every sublattice, consistent with the large-$U$ limit. 
Forming a CSL incurs an energy penalty that increases with $n$ in this strong Mott regime. 
As the Hubbard $U$ decreases, however, the CSL energy is gradually lowered, 
causing a reshuffling of energy levels within a very narrow interval, 
${4.146 \lesssim U/|t_1| \lesssim 4.161}$. 
Within this window designated as hybrid regime, 
phases with $n = 1$ to $4$ CSL successively become the ground state. 
Upon further reducing $U/|t_1|$ below $4.146$ (above ${U_c/|t_1| = 2.24}$), 
the hierarchy of low-energy states is fully inverted and the 4-CSL phase, 
features a uniform CSL on all sublattices, prevails in energy. 

The total Chern number in the $n$-CSL phase is the sum of contributions 
from all four sublattices. Since each sublattice CSL contributes $\pm 1$ 
according to the flux sign, the total Chern number can take any integer value 
between $-n$ and $+n$ in steps of $2$. For instance, the $4$-CSL phase 
can exhibit total Chern number $\pm 4$, $\pm 2$, or $0$, 
whereas the $4$-dimer ($0$-CSL) phase always has a Chern number of zero. 
It is expected that specific combinations of sublattice flux will be selected 
by higher-order corrections beyond the mean-field framework, 
ultimately determining the real Chern number of the physical ground state.

\subsection*{Photoelectric effect of CSL}

The surface CSLs realized in the weak Mott regime facilitates its detection via 
the photoelectric effect. To obtain the single-particle spectral function 
pertinent to photoemission, it is convenient to quantize the rotor operator 
into holons $a_i$ ($a_i^{\dagger}$) and doublons $b_i$ ($b_i^{\dagger}$). 
Then the electron operator is decomposed as 
\begin{equation}
       c_{\boldsymbol{k},\sigma,m}^{\dagger} = \sum_{\boldsymbol{q}} f_{\sigma,\boldsymbol{k}-\boldsymbol{q},m}^{\dagger}(a_{-\boldsymbol{q},m} + b_{\boldsymbol{q},m}^{\dagger}).
\end{equation}
The electronic Matsubara Green's function is a convolution of spinon 
and holon/doublon Green's functions 
\begin{equation}
    G_{mn} (\imath \omega, \boldsymbol{k}) = - \frac{1}{\beta} \sum_{\nu\boldsymbol{q}} G_{s,mn}(\imath\omega-\imath\nu,\boldsymbol{k}-\boldsymbol{q}) G_{r,mn}(\imath\nu,\boldsymbol{q}),
\end{equation}
where $m$ and $n$ are sublattice index. 
After finishing the summation over the Matsubara frequency $\nu$ 
for bosonic holon/doublon and the analytic continuation, 
one can obtain the retarded Green's function $G_{mn}^{R}(\omega,\boldsymbol{k})$ 
and the electron spectral function
\begin{equation}
    A(\omega \leq 0,\boldsymbol{k}) = - \frac{1}{\pi} \mathrm{Tr}\, \mathrm{Im}\, G^{R}(\omega,\boldsymbol{k}).
\end{equation}

We focus on the 4-CSL phase with the $4\boldsymbol{a}_1 \times 4\boldsymbol{a}_2$ 
enlarged unit-cell and calculate $A(\omega \leq 0,\boldsymbol{k})$ for one of four equivalent sublattices. 
The spinon and charge boson dispersions are 
\begin{equation}
    \xi_{\boldsymbol{k}}^{\pm} = \pm |\Delta| \zeta_{\boldsymbol{k}} / \sqrt{2}
\end{equation} 
and 
\begin{equation}
    \varepsilon_{\boldsymbol{k}}^{\pm} = [2U(\lambda\pm\sqrt{2}|\chi|\zeta_{\boldsymbol{k}})]^{1/2},
\end{equation} 
respectively, where
\begin{equation}
    \zeta_{\boldsymbol{k}} = [3+\cos(4k_x)-2\cos(2k_x)\cos(2\sqrt{3}k_y)]^{1/2}.
\end{equation} 
As the charge bosons experience the gauge flux generated by CSL, their dispersion also manifests two branches, which is quite different from the scenario for spinon Fermi surface in Ref.~\cite{PhysRevB.87.045119}. 
At zero temperature, the electron spectral function from the two branches of charge bosons $\varepsilon_{\boldsymbol{k}}^{\pm}$ are modulated by the dispersion of the occupied spinon band $\xi_{\boldsymbol{k}}^{-}$ as shown in Fig.~\ref{fig:monolayer}(d) for $U/|t_1| = 4.0$.
Given the large energy gap between $\varepsilon_{\boldsymbol{k}}^{\pm}$ compared with the band width of $\xi_{\boldsymbol{k}}^{-}$, 
the spectral weight is clearly distributed within two energy windows 
$\omega \approx -3.6 |t_1|$ and $\omega \approx -5.6 |t_1|$, 
corresponding to transitions involving the lower and upper charge boson bands, respectively. 
In the first window accessible with lower photon energies, 
the spectral weight concentrates near the $\overline{\mathrm{\Gamma}}$ 
(and symmetry-related) points. With increasing of photon energy, 
a similar concentration of spectral weight appears near $\overline{\mathrm{K}}$ 
(and symmetry-related) points in the second window. 
This qualitative characteristic remains distinguishable even at finite temperatures as shown in Fig.~S3 of \textit{SI Appendix}, 
and thus can serve as experimental evidence of surface CSLs. 

For the full-dimer phase ($n = 0$), it is expected that an ordered configuration shown in Fig.~\ref{fig:monolayer}(b) will be stabilized by effects beyond the mean-field framework. 
In this case, the dispersions for both spinons $\xi_{\boldsymbol{k}}^{\pm}$ and charge bosons $\varepsilon_{\boldsymbol{k}}^{\pm}$ become constants. 
Consequently, the convolved electron spectral function $A(\omega,\boldsymbol{k})$ exhibits flat bands instead of continua in the reduced Brillouin zone. 
To clearly distinguish this topologically trivial phase from the full-CSL phase ($n=4$), the temperature-dependent evolution of their electronic spectral functions $A(\omega,\boldsymbol{k})$ under identical parameters is compared in Fig.~S3 of \textit{SI Appendix}. 
Moreover, even in the hybrid phases where CSL states are established on partial sublattices, the spectral weights of the dimers remain more concentrated and dominant.
As shown in Fig.~S4 of \textit{SI Appendix}, the spectral weights of continua originating from CSLs gradually become more apparent as $n$ increases. 
However, their characteristics are only identifiable in the 4-CSL phase, where the dimer contribution disappears entirely. 
Therefore, the distinct spectral features shown in Fig.~\ref{fig:monolayer}(e) allow the topologically non-trivial 4-CSL phase to be experimentally identified and separated from other surface states via ARPES.

\section{Discussion}

The surface CSL 
offers a direct microscopic mechanism for the field-free Josephson diode effect 
observed in NbSe$_2$/Nb$_3$X$_8$/NbSe$_2$ (X = Br, Cl) vdW 
junctions~\cite{Wu2022,dubbelman2025drivingfieldfreejosephsondiode}.
The persistence of the nonreciprocal supercurrent down to the monolayer 
limit~\cite{doi:10.1021/acs.nanolett.5c04336}, without an external magnetic field, 
necessitates a spontaneous time-reversal symmetry breaking originating from the barrier itself. 
Critically, growing experimental evidence has established the non-magnetic Mottness 
in Nb$_3$X$_8$ compounds~\cite{doi:10.1021/acsnano.9b04392,PhysRevX.13.041049,mzyy-thjq,10.1016/j.newton.2025.100292,Hu2023,wr7w-nfhg}.
This leads naturally to the CSL phase we identify, 
whose intrinsic chiral Abelian topological order emerging from 
the electron corrections provides the requisite symmetry breaking.
A similar SSH stacking has been analyzed in multilayer 1T-TaS$_2$ cluster
Mott insulator~\cite{PhysRevLett.129.017202} where the type-II surface was proposed 
to realize the spinon Fermi surface instead of CSL. Since both systems were understood
from the single-band Hubbard model in the weak Mott regime~\cite{PhysRevLett.121.046401,Persky2022,PhysRevLett.127.087201,Luo2024,mzyy-thjq,PhysRevX.13.041049}, it would be interesting if one can
apply pressures or chemical pressure to tune the system through different surface states.

The predicted ARPES signature of the surface CSL, spectral weight concentration 
near high-symmetry points in distinct energy windows, offers a clear experimental fingerprint. 
Beyond photoemission, other surface-sensitive probes could further test our proposal, 
such as scanning tunnelling microscopy for real-space modulations, 
magneto-optical Kerr effect for surface chirality, and microwave impedance microscopy 
for local compressibility. Transport measurements on the exfoliated flakes with controlled terminations 
could also reveal quantized thermal Hall conductance associated with the chiral edge modes of the CSL.
On the theoretical side, future work could explore the stability of the CSL phase against disorder,
inter-sublattice couplings, and deviations from perfect half-filling.
It would be valuable to study the dynamical properties of the surface topological order, 
such as its characteristic spin and charge response functions, 
which could be compared with future dynamical probes. 
Finally, the interplay between surface topological order and superconductivity, 
as hinted by the Josephson diode experiments,
opens a fascinating direction for exploring superconductivity, flux, exotic fractional excitations 
in fully van der Waals-integrated architectures.

In summary, we have elucidated a possible realization of ITO on the type-II terminated surface of Nb$_3$Br$_8$, 
by combining first-principles calculations with a constrained slave-rotor mean-field theory. 
Furthermore, we predict its characteristic ARPES signature, providing a concrete path for experimental detection.
Our work not only proposes specific material realization of surface topological order but also highlights a broader conceptual framework for discovering and controlling correlated topological phases in layered quantum materials.

\section{Materials and methods}

The first-principles calculations for the electronic structure of Nb$_3$Br$_8$ are performed in the framework of Kohn-Sham density functional theory~\cite{Hohenberg1964,Kohn1965}, as implemented in the \textsc{Quantum} ESPRESSO package~\cite{Giannozzi2009,Giannozzi2017}. The projector augmented-wave pseudopotentials in the pslibrary~\cite{DalCorso2014,pslibrary} were adopted along with Perdew-Burke-Ernzerhof exchange-correlation functional~\cite{Perdew1996}. The crystal structure in Ref.~\cite{Simon1966} was used in the calculations. A cutoff energy of 140~Ry was chosen for the plane wave basis set. The Brillouin zone was sampled by a $ 3\times3\times3 $ Monkhorst-Pack mesh. The maximally localized Wannier functions were constructed by the Wannier90 package~\cite{Marzari1997,Pizzi2020}, and the low-energy tight-binding model was thus obtained. To calculate the surface spectral function, we implemented the iterative procedure in Ref.~\cite{Sancho1985} to get the surface Green's function.
The band structure of bulk Nb$_3$Br$_8$ and the high-symmetry points and lines in the Brillouin zone are shown in the \textit{SI Appendix}.

\subsection{Self-consistent slave-rotor mean-field theory}

For the type-II terminated surface of Nb$_3$Br$_8$, the relevant electronic structure is described by a monolayer triangular lattice, which is weakly coupled to the bulk via interlayer hopping with a ratio $t_{\perp'}/t_{\perp} \approx 0.118$ from the DFT calculations.  
Within the slave-rotor framework, it is therefore justified to focus on the intralayer electron hopping within this surface monolayer.  
Based on our DFT simulations, we set the corresponding hopping parameters in the Hamiltonian~\eqref{eq:hubbard} to  $t_1 = -1$, $t_2/|t_1| = -1.09$, and $t_3/|t_1| = +1.37$. 
To comprehensively search for different saddle-point solutions, the unit cell of the pristine triangular lattice is enlarged along $\boldsymbol{a}_{1}$ and $\boldsymbol{a}_2$ directions by factors of $\ell_1$ and $\ell_2$, respectively. 
Under periodic boundary conditions, the reduced Brillouin zone is simultaneously discretized into an $L_1 \times L_2$ mesh. 
In our numerical iteration process, we explore the enlargement factor $\ell_{1,2}$ ranging from 2 to 6 with a fixed $L_{1,2} = 30$. 
Once the convergence to the saddle-point solution is achieved, the results are further refined using a large $L_{1,2} = 100$ to ensure high precision.
For each value of $U$, a total of 1024 different configurations were initialized with random $\Delta_{ij}$ and $\chi_{ij}$ on the enlarged unit cell.
All results that achieved energy convergence were retained, after filtering out rare instances where convergence was reached despite violations of the local spinon or global rotor constraints. 
The detailed derivations of the slave rotor mean-field theory for the half-filled Hubbard Hamiltonian and the implemention of the self-consistent minimization algorithm are provided in the \textit{SI Appendix}.

\subsection{Calculation of Chern numbers}
For a given enlarged unit cell, once the self-consisitent minimization algorithm finds a saddle-point solution, the spinon mean-field Hamiltonian $H_s$ in~\eqref{eq:MF_spinon} becomes physically well defined.
For both the $n$-CSL and dimer states, a finite energy gap separates the occupied and unoccupied spinon bands. 
However, due to the U(1) gauge pattern and enlargement of unit cell, the conduction spinon bands exhibite many crossings in the reduced Brillouin zone. 
To determine the topological invariant of occupied spinon states, we employ the generalized form of the first Chern number~\cite{PhysRevB.105.024401}
\begin{equation}
    C_1 = \frac{1}{2\pi} \int_{\boldsymbol{k}\in\text{BZ}} d^2\boldsymbol{k} \mathrm{Tr}[\mathcal{F}_{ij}(\boldsymbol{k})].
\end{equation}
Here, the Berry curvature $\mathcal{F}_{ij} = \frac{\partial \mathcal{A}_j}{\partial k_i} - \frac{\partial \mathcal{A}_i}{\partial k_j} - \imath [\mathcal{A}_i, \mathcal{A}_j]$ is non-Abelian and expressed in terms of the matrix-valued Berry connection $\mathcal{A}_i$. 
The matrix elements are $[\mathcal{A}_i]_{nm} = \imath \langle \psi_n | \frac{\partial}{\partial k_i} | \psi_m \rangle$, where $\psi_n(\boldsymbol{k})$ denotes the $n$-th spinon eigenstate from $\mathcal{U}_{\boldsymbol{k}}$ and $\boldsymbol{k} = (k_i, k_j)$ is the momentum in the reduced Brillouin zone.  
For each of the four sublattices, the occupied spinon bands possess $C_1 = 0$ in the dimer state and $C_1 = \pm 1$ in the CSL state. 
In the latter case, the sign of the first Chern number reflects the chirality of the U(1) gauge flux, and is obtained randomly by our numerical algorithm. 
Consequently, the total first Chern number computed from the above general definition ranges from $-n$ to $+n$ in steps of 2 for every $n$-CSL phases when starting from different random seeds.

\subsection{Intertwinded electron Green's function} 

Under the slave-rotor framework, the the partition function for the Hubbard model can be written as 
\begin{equation}
    \mathcal{Z}=\int\mathcal{D}[f^{\dagger}, f, \Phi^{*}, \Phi, h, \lambda] e^{-\mathcal{S}_s-\mathcal{S}_c}.
\end{equation}
The explicit forms of $\mathcal{S}_s$ and $\mathcal{S}_c$ are provided in \textit{SI Appendix}. 
The rotor operators can be further quantized into charge bosons by defining the canonical momenta 
\begin{align}
    \Pi_i & = (\partial_t - \imath h_i)\Phi_i/(2U),\\
    \Pi_i^{\dagger} & = (\partial_t + \imath h_i)\Phi_i^{*}/(2U), 
\end{align}
which respect the commutation relations $[\Phi_i, \Phi_j^{*}] = [\Pi_i, \Pi_j^{\dagger}]=0$, and $[\Phi_i, \Pi_j] = [\Phi_i^{*}, \Pi_j^{\dagger}]= \imath\delta_{ij}$. 
Following the standard canonical quantization procedure, two types of bosonic operators are introduced~\cite{PhysRevB.107.195155}
\begin{align}
    a_{i}^{(\dagger)} & = \frac{1}{\sqrt{2}}\left[\left(\frac{\lambda}{2U}\right)^{1/4}\Phi_i + \imath\left(\frac{\lambda}{2U}\right)^{-1/4}\Pi_i\right]^{(\dagger)}, \label{eq:doublon}\\
    b_{i}^{(\dagger)} & = \frac{1}{\sqrt{2}}\left[ \left(\frac{\lambda}{2U}\right)^{1/4}\Phi_i^{*} + \imath\left(\frac{\lambda}{2U}\right)^{-1/4}\Pi_i^{\dagger} \right]^{(\dagger)}. \label{eq:holon}
\end{align}
Physically, these bosonic operators annihilate (create) a holon and a doublon at site $i$, respectively. 
In this basis, the mean field rotor Hamiltonian takes a bosonic Bogoliubov-de-Gennes form
\begin{equation}
    H_r = \sqrt{2U\lambda}  \sum_i a_i a_i^{\dagger}+b_i^{\dagger}b_i 
    +  \sum_{ij}
    \begin{pmatrix}
        a_i & b_i^{\dagger}
    \end{pmatrix}
    \mathcal{M}
    \begin{pmatrix}
        a_j^{\dagger} \\
        b_j
    \end{pmatrix} + \text{h.c.}
\end{equation}
where
\begin{equation}
    \mathcal{M}=\sqrt{U/(2\lambda)}\chi_{ij}
    \begin{pmatrix}
        1 & 1 \\
        1 & 1
    \end{pmatrix}.
\end{equation}
In the momentum space, this Bogoliubov-de-Gennes Hamiltonian can be diagonalized as
\begin{equation}
    H_r = \sum_{\boldsymbol{k},n} \varepsilon_{\boldsymbol{k},n} \beta_{\boldsymbol{k},n}^{\dagger} \beta_{\boldsymbol{k},n} + \varepsilon_{-\boldsymbol{k},n} \alpha_{-\boldsymbol{k},n} \alpha_{-\boldsymbol{k},n}^{\dagger},
\end{equation}
via a paraunitary transformation
\begin{equation}
    \mathcal{P}_{\boldsymbol{k}} = 
    \begin{pmatrix}
        \boldsymbol{A}(\boldsymbol{k}) & \boldsymbol{B}(\boldsymbol{k}) \\
        \boldsymbol{C}(\boldsymbol{k}) & \boldsymbol{D}(\boldsymbol{k})
    \end{pmatrix}.
\end{equation} 
Therefore, the bosonic Green's function for holon and doublons takes the following form 
\begin{equation}
    G_{r,mn}(\imath\nu,\boldsymbol{q}) = \sum_{l} \frac{c_{mn,l}}{\imath\nu-\varepsilon_{\boldsymbol{q},l}} + \frac{d_{mn,l}}{-\imath\nu-\varepsilon_{-\boldsymbol{q},l}},
\end{equation}
where $c_{mn,l} = (\boldsymbol{C}_{\boldsymbol{q},ml}+\boldsymbol{A}_{\boldsymbol{q},ml})(\boldsymbol{C}_{\boldsymbol{q},ln}^{\dagger}+\boldsymbol{A}_{\boldsymbol{q},ln}^{\dagger})$ and $d_{nm,l} = (\boldsymbol{D}_{\boldsymbol{q},ml}+\boldsymbol{B}_{\boldsymbol{q},ml})(\boldsymbol{D}_{\boldsymbol{q},ln}^{\dagger}+\boldsymbol{B}_{\boldsymbol{q},ln}^{\dagger})$.
The subscripts indicate different sublattices and $\varepsilon_{\boldsymbol{k}}$ are eigenvalues afther the paraunitary transformation. 
The fermionic Green's function for spinons is 
\begin{equation}
   G_{s,mn} (\imath\omega, \boldsymbol{k}-\boldsymbol{q}) = \sum_{l} \frac{\mathcal{U}_{\boldsymbol{k}-\boldsymbol{q},ml} \mathcal{U}_{\boldsymbol{k}-\boldsymbol{q},ln}^{\dagger}}{\imath\omega - \xi_{\boldsymbol{k}-\boldsymbol{q},l}},
\end{equation}
where $\mathcal{U}(\boldsymbol{k})$ is the unitary matrix that diagonalizes the spinon mean-field Hamiltonian $H_s(\boldsymbol{k})$. 
After finishingthe summation over the Matsubara frequency and the analytic continuation, the retarded Green's funtion for electrons is 
\begin{align}
    G_{mn}^{R} (\omega,\boldsymbol{k}) = \sum_{\boldsymbol{q},ll'} & \frac{c'_{mn,ll'}}{\omega+\imath 0^+-\xi_{\boldsymbol{k}-\boldsymbol{q},l}-\varepsilon_{+\boldsymbol{q},l'}} \notag \\
    + & \frac{d'_{mn,ll'}}{\omega+\imath 0^+-\xi_{\boldsymbol{k}-\boldsymbol{q},l}+\varepsilon_{-\boldsymbol{q},l'}},
\end{align}
where $c'_{mn,ll'} = c_{mn,l'} \mathcal{U}_{\boldsymbol{k}-\boldsymbol{q},ml} \mathcal{U}_{\boldsymbol{k}-\boldsymbol{q},ln}^{\dagger} [n_F(-\xi_{\boldsymbol{k}-\boldsymbol{q},l}) + n_B(\varepsilon_{\boldsymbol{q},l'})]$ and $d'_{mn,ll'} = d_{mn,l'} \mathcal{U}_{\boldsymbol{k}-\boldsymbol{q},ml} \mathcal{U}_{\boldsymbol{k}-\boldsymbol{q},ln}^{\dagger} [n_F(\xi_{\boldsymbol{k}-\boldsymbol{q},l}) + n_B(\varepsilon_{-\boldsymbol{q},l'})]$.
$n_F(\xi)=1/(e^{\beta\xi}+1)$ and $n_B(\varepsilon)=1/(e^{\beta\varepsilon}-1)$ are the Fermi and Boson distribution function, respectively. 
The detailed derivations of the intertwinded electron Green's functions are provided in \textit{SI Appendix}.

\subsection{Electron spectral function of surface CSL}

The electron spectral functions are calculated for the surface 4-CSL phase where the same CSL is realized in all four sublattices, corresponding to the maximal total Chern number $\pm 4$. 
Becuase of the $\pm \pi/2$ flux on each triangular plaquette formed by the auxiliary field on third nearest neighbor bonds, the unit cell on each sublattice must be doubled along one of the sublattice vectors, as shown in Fig.~\ref{fig:monolayer}(a). 
Nevertheless, the minimal enlarged unit cell required to capture the full $4$-CSL state is of size $4\boldsymbol{a}_1 \times 4\boldsymbol{a}_2$, due to the sublattice structure, which consists of four sites per sublattice. 
The reduced Brillouin zone and the the high-symmetry points are shown in \textit{SI Appendix}. 
By convolving the Green's functions corresponding to $\xi_{\boldsymbol{k}}^{\pm}$ and $\varepsilon_{\boldsymbol{k}}^{\pm}$, the Lorentzian broadening parameter is set to $0.01$ without loss of any generality. 
The zero-temperature result, as presented in Fig.~\ref{fig:monolayer}(e) is appropriated by setting $\beta = 100/|t_1|$ for $U/|t_1| = 4.0$. 
The numerical details and the electron spectral functions at various temperatures and for other surface phases are provided in the \textit{SI Appendix}.

\section{Acknowledgments}
We acknowledge Rui Luo for a previous collaboration. 

\section{Supplementary material}
Supplementary material is available at \textit{PNAS Nexus} online.

\section{Competing interest}
The authors declare that they have no competing interests.

\section{Funding}
C.-K. Li acknowledges the support of the NSFC No.~12404126 and the Start-up Research Fund of Southeast University (Grant No.~RF1028624145).
G.C. is supported by Quantum Science and Technology-National Science and Technology Major Project (grant No.~2025ZD0300500), and by NSFC No.~12574061 and No.~92565110.

\section{Author contributions}

X.-P.Y. and C.-K.L. performed the analytical and numerical calculations. 
G.C. initiated and supervised the project. All authors contributed to theoretical analysis and the writing of the manuscript.

\section{Preprints}
This manuscript was posted on a preprint: \\
https://doi.org/10.48550/arXiv.2601.05185.

\section{Data availability}

All study data are included in the article and/or \textit{SI Appendix}.

% \section{Ethics statement}
% This study utilized data from the XXX study, which obtained
% ethics approval from the YYY Committee (approval number: ZZZ), and obtained written informed consent from all participants prior to the study in accordance with the Declaration of
% Helsinki. 

%IF YOU WISH TO INCLUDE A BIB FILE TO BUILD THE REFERENCE LIST, UNCOMMENT THE BELOW TWO LINES AND REMOVE THE {thebibliography} EXAMPLE THAT FOLLOWS THEM.
\bibliographystyle{unsrt}
\bibliography{Ref}

\end{document}

\title{Surface chiral Abelian topological order on multilayer cluster Mott insulators}

\author{Xu-Ping Yao}
\thanks{These authors contributed equally.}
\affiliation{Kavli Institute for Theoretical Sciences, University of Chinese Academy of Sciences, Beijing 100190, China}

\author{Chao-Kai Li}
\thanks{These authors contributed equally.}
\affiliation{School of Physics, Southeast University, Nanjing 211189, China}

\author{Gang v.~Chen}
\email{chenxray@pku.edu.cn}
\affiliation{International Center for Quantum Materials, School of Physics, Peking University, Beijing 100871, China}
\affiliation{Collaborative Innovation Center of Quantum Matter, 100871, Beijing, China}

\date{\today}

% \begin{abstract}

% \end{abstract}

\maketitle

\subsection*{First-principles calculations}

\begin{figure}[h]
    \centering
	\includegraphics[width=0.75\textwidth]{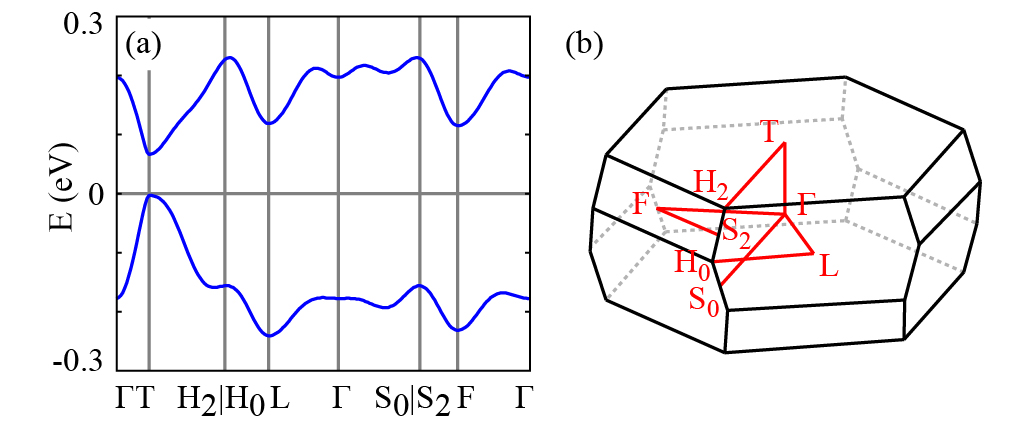}
	\caption{
	(a) The bulk band structure of Nb$_3$Br$_8$. (b) The Brillouin zone of bulk Nb$_3$Br$_8$.}
	\label{fig: band_structures_bulk}
\end{figure}

In Fig.~\ref{fig: band_structures_bulk}(a) is shown the band structure of bulk Nb$_3$Br$_8$. The high-symmetry lines were found by the SeeK-path program~\cite{Hinuma2017,Togo2018}. A direct band gap of around $ 70~\text{meV} $ is located at the T point of the Brillouin zone [Fig.~\ref{fig: band_structures_bulk}(b)]. The dispersion brought about by the in-plane hoppings does not destroy the insulating nature of the SSH model along the stacking direction.

\subsection*{Self-consistent slave-rotor mean-field theory}

In the following, we detail the derivations of the slave rotor mean-field theory for the half-filled Hubbard Hamiltonian
\begin{equation}\label{eq:hubbard}
    H = - \sum_{ij,\sigma} (t_{ij} c_{i\sigma}^{\dagger}c_{j\sigma} + \text{h.c.}) + \frac{U}{2} \sum_i (n_{i} - 1)^2,
\end{equation}
where $c_{i\sigma}^{\dagger}$ and $c_{i\sigma}$ are creation annihilation operators for electrons with spin $\sigma = \uparrow$ or $\downarrow$ at site $i$. 
$n_{i} = \sum_{\sigma}c_{i\sigma}^{\dagger}c_{i\sigma}$, $t_{ij}$, and $U$ denotes the electron number, the hopping energy, and the on-site Coulomb repulsion, respectively. 
Following the procedures developed in Refs.~\cite{PhysRevB.70.035114,PhysRevLett.95.036403}, the electron operators is decomposed into 
\begin{equation}
    c_{i\sigma}^{\dagger} = f_{i\sigma}^{\dagger} \Phi_i, \quad c_{i\sigma}=f_{i\sigma} \Phi_i^{*}, 
\end{equation}
where $\Phi_i=e^{\imath\theta_i}$. 
The fermionic spinon operator $f^{\dagger}$ (and $f$) and the bosonic rotor operator $\Phi^{*}$ (and $\Phi$) carry the spin and charge degrees of freedom, respectively. 
The charge quantum number corresponds to an angular momentum $L_i = -\imath \partial_{\theta_i}$, satisfying $[\theta_i, L_j]=\imath\delta_{ij}$, and is constrained by $L_i=\sum_{\sigma}f_{i\sigma}^{\dagger}f_{i\sigma}-1$.
In this representation, the Hubbard model in Eq.~\eqref{eq:hubbard} is rewritten as 
\begin{equation}
    H = \sum_{ij,\sigma} t_{ij} e^{\imath(\theta_i-\theta_j)} f_{i\sigma}^{\dagger} f_{j\sigma} + \text{h.c} - \sum_{i\sigma} (h_i + \mu_i) f_{i\sigma}^{\dagger}f_{i\sigma} + \sum_i \frac{U}{2}L_i^2 + h_i L_i + h_i + \mu_i.
\end{equation}
We have introduced a Lagrange multiplier $h_i$ to enforce the local constraint on $L_i$. 
A field $\mu_i$ has also been incorporated as the chemical potential to reduce the enlarged local Hilbert space for spinons. 
The corresponding Euclidean action is
\begin{equation}
    \mathcal{S} = \int_{0}^{\beta} d\tau \left [ \sum_{i\sigma} f_{i\sigma}^{\dagger}(\partial_{\tau} - h_i - \mu_i) f_{i\sigma} + \frac{1}{2U} \sum_i (\partial_{\tau} \theta_i + \imath h_i)^2  - \sum_{ij,\sigma} t_{ij}f_{i\sigma}^{\dagger}f_{j\sigma}e^{\imath(\theta_i-\theta_j)} + \text{c.c.} + \sum_i (\mu_i + h_i) \right ]. 
\end{equation}
The non-quadratic hopping term in the second line, which couples spinons and rotors, can be decoupled via the Hubbard-Stratonovich transformation. 
At the saddle point, the expectations of the resulting auxiliary fields after the decomposition are defined as
\begin{equation}\label{eq:auxiliary_fields}
    \Delta_{ij} = -t_{ij} \langle \Phi_i\Phi_j^{*} \rangle, \quad \chi_{ij} = -t_{ij} \sum_{\sigma} \langle f_{i\sigma}^{*}f_{j\sigma} \rangle.
\end{equation}
This yields the following mean-field Hamiltonians for the spinon and rotor sectors
\begin{align}
    H_s & = \sum_{ij,\sigma} \Delta_{ij} f_{i\sigma}^{\dagger}f_{j\sigma} + \text{h.c.} + \sum_i \mu_i (1 - f_{i\sigma}^{\dagger}f_{i\sigma}), \label{eq:MF_spinon} \\
    H_r & = \sum_{i\sigma} \chi_{ij}^{*} \Phi_i^{*} \Phi_j + \text{h.c.} + \sum_i \frac{U}{2} L_i^2, 
\end{align}
supplemented by a constant energy shift $\sum_{ij} \Delta_{ij}\chi_{ij}/t_{ij}$. 
Note that the expectation of linear term of $-h_i L_i$ vanishes at the saddle point, allowing $h_i$ to be absorbed into $\mu_i$, as they play an identical role in the remaining mean-field Hamiltonians.

In momentum space, the spinon mean-field Hamiltonian $H_s$ is diagonalized by a unitary transformation
\begin{equation}
    \mathcal{U}^{\dagger}_{\boldsymbol{k}} H_s(\boldsymbol{k}) \mathcal{U}_{\boldsymbol{k}} = \mathrm{diag} \{ \xi_{\boldsymbol{k},1}, \xi_{\boldsymbol{k},2}, \ldots, \xi_{\boldsymbol{k},n} \}. 
\end{equation} 
To obtained the saddle-point solutions that strictly satisfy the spinon local constraints $\sum_{\sigma} f_{i\sigma}^{\dagger}f_{i\sigma} = 1$, we employ the self-consistent minimization algorithm developed in Refs.~\cite{PhysRevLett.103.135301,PhysRevB.84.174441} for spinons and its extensions in Refs.~\cite{PhysRevA.93.061601,PhysRevB.105.024401} with sublattice and rotor degree of freedom. 
Once a solution satisfying the spinon local constraints is found, it gives an auxiliary field $\chi_{ij}$ via the spinon correlator $\langle f_{i\sigma}^{\dagger}f_{j\sigma} \rangle$, which in turn specifies the rotor Hamiltonian $H_r$.  
For the rotor Hamiltonian $H_r$, the unimodular condition $\Phi_i^{*}\Phi_i  = |\Phi_i|^2 = 1$ introduces an implicit local constraint, enforced by another Lagrange multiplier $\lambda_i$. 
We adopt a uniform saddle-point approximation in the rotor sector by setting $\lambda_i = \lambda$. 
After the Fourier transformation in both momentum and frequency space, the effective rotor action becomes 
\begin{equation}
    \mathcal{S}_r = \int_{0}^{\beta} d\tau \sum_{\boldsymbol{k}} \left[ \frac{1}{2U} | \partial_{\tau} \Phi_{\boldsymbol{k}} |^2  - \lambda + (\lambda + \sum_{\boldsymbol{d}} \chi_{ij}^{*} e^{\imath \boldsymbol{d}_{ij}\cdot\boldsymbol{k}} + \text{h.c.}) |\Phi_{\boldsymbol{k}}|^2 \right],
\end{equation}
where $\boldsymbol{d}=\boldsymbol{r}_j-\boldsymbol{r}_i$ is the displacement vector connecting sites $i$ and $j$.
From this action, the rotor correlator is evaluated as
\begin{equation}
    \langle \Phi_{\boldsymbol{k}\alpha}^{*}\Phi_{\boldsymbol{k}\beta} \rangle = \sum_{n} \mathcal{V}_{\boldsymbol{k}\alpha n}^{*}\mathcal{V}_{\boldsymbol{k}\beta n} \frac{U}{\sqrt{2U(\epsilon_{\boldsymbol{k} n} + \lambda)}},
\end{equation}
where $n$, $\alpha$, and $\beta$ index the sublattice bands.  
$\epsilon_{\boldsymbol{k}}$ and $\mathcal{V}_{\boldsymbol{k}}$ are eigenvalues and transformation matrices that diagonalize the the quadratic part of $\mathcal{S}_r$. 
The self-consistent equation enforcing the relaxed global constraint over all $N_s$ sites is then 
\begin{equation}
    \frac{1}{N_s} \int_{\boldsymbol{k}\in\text{BZ}} d^2\boldsymbol{k} \sum_{n} \langle \Phi_{\boldsymbol{k} n}^{*}\Phi_{\boldsymbol{k} n} \rangle = 1.
\end{equation}
The rotor correlator updates the auxiliary field $\Delta_{ij}$ as well after the inverse Fourier transformation, thereby completing the self-consistent procedure between both sectors. 

\subsection*{The photoelectric effect of surface chiral spin liquids}

\subsubsection*{The intertwinded electron Green's function}

Following the slave-rotor description, the partition function for the Hubbard model can be written as 
\begin{equation}
    \mathcal{Z}=\int\mathcal{D}[f^{\dagger}, f, \Phi^{*}, \Phi, h, \lambda] e^{-\mathcal{S}_s-\mathcal{S}_c},
\end{equation}
where spinon action $\mathcal{S}_s$ and rotor action $\mathcal{S}_r$ are  
\begin{align}
    \mathcal{S}_s & = \int_{0}^{\beta} d\tau \left[ \sum_{i\sigma}f_{i\sigma}^{\dagger}(\partial_{\tau} -h_i - \mu_i) f_{i\sigma} + \sum_{ij,\sigma} \Delta_{ij} f_{i\sigma}^{\dagger}f_{i\sigma} + \text{h.c.} + \sum_i(\mu_i + h_i) \right], \\
    \mathcal{S}_r & = \int_{0}^{\beta} d\tau \left[ \frac{1}{2U}\sum_i [(\partial_{\tau} + h_i)\Phi_i^{*}][(\partial_{\tau} - h_i)\Phi_i] + \sum_{ij} \chi_{ij}^{*} \Phi_i^{*} \Phi_j + \text{h.c.} + \lambda\sum_{i} (|\Phi_i|^2-1) \right],
\end{align}
To calculate the single-particle spectral function pertinent to photoemission, it is convenient to quantize the rotor operators into a quasiparticle picture. 
This is achieved by defining the canonical momenta from the Euclidean Lagrangian after a Wick rotation 
\begin{equation}
    \Pi_i = (\partial_t - \imath h_i)\Phi_i/(2U), \quad \Pi_i^{\dagger} = (\partial_t + \imath h_i)\Phi_i^{*}/(2U), 
\end{equation}
which respect the commutation relations $[\Phi_i, \Phi_j^{*}] = [\Pi_i, \Pi_j^{\dagger}]=0$, and $[\Phi_i, \Pi_j] = [\Phi_i^{*}, \Pi_j^{\dagger}]= \imath\delta_{ij}$. 
The rotor Hamiltonian in terms of $\Pi$ and $\Phi$ operators then reads  
\begin{equation}
    H_r = 2U \Pi_i^{\dagger}\Pi_i + \imath h_i(\Pi_i^{\dagger}\Phi_i^{*} - \Pi_i \Phi_i) + \lambda \sum_{i} \Phi_i^{\dagger}\Phi_i + \sum_{ij} [\chi_{ij}^{*} \Phi_i^{*}\Phi_j + \text{h.c.}].
\end{equation}
Following the standard canonical quantization procedure, two types of bosonic operators are introduced~\cite{PhysRevB.107.195155}
\begin{align}
    a_{i}^{(\dagger)} & = \frac{1}{\sqrt{2}}\left[\left(\frac{\lambda}{2U}\right)^{1/4}\Phi_i + \imath\left(\frac{\lambda}{2U}\right)^{-1/4}\Pi_i\right]^{(\dagger)}, \label{eq:doublon}\\
    b_{i}^{(\dagger)} & = \frac{1}{\sqrt{2}}\left[ \left(\frac{\lambda}{2U}\right)^{1/4}\Phi_i^{*} + \imath\left(\frac{\lambda}{2U}\right)^{-1/4}\Pi_i^{\dagger} \right]^{(\dagger)}. \label{eq:holon}
\end{align}
Physically, these bosonic operators annihilate (create) a holon and a doublon at site $i$, respectively. 
In this basis, the mean field rotor Hamiltonian takes a bosonic Bogoliubov-de-Gennes (BdG) form 
\begin{equation}
    H_r = \sqrt{2U\lambda}  \sum_i a_i a_i^{\dagger}+b_i^{\dagger}b_i 
    +  \sum_{ij}
    \begin{pmatrix}
        a_i & b_i^{\dagger}
    \end{pmatrix}
    \mathcal{M}
    \begin{pmatrix}
        a_j^{\dagger} \\
        b_j
    \end{pmatrix} + \text{h.c.}
\end{equation}
where
\begin{equation}
    \mathcal{M}=\sqrt{U/(2\lambda)}\chi_{ij}
    \begin{pmatrix}
        1 & 1 \\
        1 & 1
    \end{pmatrix}.
\end{equation}
As noted previously, the Lagrangian multiplier $h_i$ can be absorbed into $\mu_i$ in the spinon sector and is henceforth neglected. 
Performing a Fourier transformation yields the momentum-space rotor Hamiltonian
\begin{equation}
    H_{r} = \sum_{\boldsymbol{k}}
    \begin{pmatrix}
        b_{+\boldsymbol{k}}^{\dagger}, a_{-\boldsymbol{k}}
    \end{pmatrix}
    \mathcal{M}(\boldsymbol{k})
    \begin{pmatrix}
        b_{+\boldsymbol{k}} \\
        a_{-\boldsymbol{k}}^{\dagger}
    \end{pmatrix},
\end{equation}
where 
\begin{equation}
    \mathcal{M}(\boldsymbol{k}) = \sqrt{U/(2\lambda)} \sum_{\boldsymbol{d}}
    [\chi_{ij}^{*} e^{\imath\boldsymbol{d}_{ij}\cdot\boldsymbol{k}} + \text{c.c.}]
    \begin{pmatrix}
        1 & 1 \\
        1 & 1
    \end{pmatrix}
    + \sqrt{2U\lambda} \mathbb{I},
\end{equation}
and $\mathbb{I}$ is the identity matrix. 
The bosonic BdG Hamiltonian is diagonalized as
\begin{equation}
    H_r = \sum_{\boldsymbol{k},n} \varepsilon_{\boldsymbol{k},n} \beta_{\boldsymbol{k},n}^{\dagger} \beta_{\boldsymbol{k},n} + \varepsilon_{-\boldsymbol{k},n} \alpha_{-\boldsymbol{k},n} \alpha_{-\boldsymbol{k},n}^{\dagger},
\end{equation}
via a paraunitary transformation~\cite{PhysRevB.87.174427,PhysRevB.100.075414} 
\begin{equation}
    \mathcal{P}_{\boldsymbol{k}} = 
    \begin{pmatrix}
        \boldsymbol{A}(\boldsymbol{k}) & \boldsymbol{B}(\boldsymbol{k}) \\
        \boldsymbol{C}(\boldsymbol{k}) & \boldsymbol{D}(\boldsymbol{k})
    \end{pmatrix},
\end{equation}
under which the new basis $(\boldsymbol{\beta}_{\boldsymbol{k}}^{\dagger},\boldsymbol{\alpha}_{-\boldsymbol{k}})$ relates to the old one by $(\boldsymbol{b}_{\boldsymbol{k}}^{\dagger},\boldsymbol{a}_{-\boldsymbol{k}})\mathcal{P}_{\boldsymbol{k}} = (\boldsymbol{\beta}_{\boldsymbol{k}}^{\dagger},\boldsymbol{\alpha}_{-\boldsymbol{k}})$, with $\boldsymbol{b}_{\boldsymbol{k}}^{\dagger} = (b_{\boldsymbol{k},1}^{\dagger},b_{\boldsymbol{k},2}^{\dagger},\ldots,b_{\boldsymbol{k},n}^{\dagger})$ and $\boldsymbol{a}_{-\boldsymbol{k}}=(a_{-\boldsymbol{k},1},a_{-\boldsymbol{k},2},\ldots,a_{-\boldsymbol{k},n})$ for $n$ sublattices.

In terms of spinons and holons/doublons, the physical electron operator is decomposed as 
\begin{equation}
    c_{\boldsymbol{k},\sigma,m}^{\dagger} = \sum_{\boldsymbol{q}} f_{\sigma,\boldsymbol{k}-\boldsymbol{q},m}^{\dagger}(a_{-\boldsymbol{q},m} + b_{\boldsymbol{q},m}^{\dagger}),
\end{equation} 
according to Eqs.~\eqref{eq:doublon} and~\eqref{eq:holon}. 
The spin index $\sigma$ in the spinon sector is suppressed hereafter for clarity. 
It is straightforward to express the electronic Matsubara Green's function as a convolution of spinon and holon/doublon Green's functions
\begin{equation}
    G_{mn} (\imath \omega, \boldsymbol{k}) = - \int_{0}^{\beta} e^{\imath\omega\tau} d\tau \langle c_{\boldsymbol{k},n}(\tau) c_{\boldsymbol{k},m}^{\dagger}(0) \rangle 
    = - \frac{1}{\beta} \sum_{\nu\boldsymbol{q}} G_{s,mn}(\imath\omega-\imath\nu,\boldsymbol{k}-\boldsymbol{q}) G_{r,mn}(\imath\nu,\boldsymbol{q}).
\end{equation}
The fermionic spinon Green's function follows directly from the unitary diagonalization
\begin{align}
    G_{s,mn} (\imath\omega, \boldsymbol{k}-\boldsymbol{q}) & = \sum_{l} \frac{\mathcal{U}_{\boldsymbol{k}-\boldsymbol{q},ml} \mathcal{U}_{\boldsymbol{k}-\boldsymbol{q},ln}^{\dagger}}{\imath\omega - \xi_{\boldsymbol{k}-\boldsymbol{q},l}}.
\end{align}
The bosonic Green's function for holon and doublons is more complicated due to the BdG structure 
\begin{equation}
    G_{r,mn}(\imath\nu,\boldsymbol{q}) = \sum_{l} \frac{c_{mn,l}}{\imath\nu-\varepsilon_{\boldsymbol{q},l}} + \frac{d_{mn,l}}{-\imath\nu-\varepsilon_{-\boldsymbol{q},l}}
\end{equation}
where 
\begin{align}
    c_{mn,l} & = (\boldsymbol{C}_{\boldsymbol{q},ml}+\boldsymbol{A}_{\boldsymbol{q},ml})(\boldsymbol{C}_{\boldsymbol{q},ln}^{\dagger}+\boldsymbol{A}_{\boldsymbol{q},ln}^{\dagger}), \\
    d_{nm,l} & = (\boldsymbol{D}_{\boldsymbol{q},ml}+\boldsymbol{B}_{\boldsymbol{q},ml})(\boldsymbol{D}_{\boldsymbol{q},ln}^{\dagger}+\boldsymbol{B}_{\boldsymbol{q},ln}^{\dagger}).
\end{align}
We have used the Green's funtions $G_{\beta, l}(\imath\nu,\boldsymbol{q}) = 1/(\imath\nu-\varepsilon_{\boldsymbol{k},l})$ and $G_{\alpha,l}(\imath\nu,\boldsymbol{q}) = 1/(-\imath\nu-\varepsilon_{-\boldsymbol{q},l})$ in the diagonalized basis.
After finishing the summation over the Matsubara frequency $\nu$ for bosonic holon/doublon, the electronic Green's function can be obtained 
\begin{equation}
    G_{mn} (\imath\omega,\boldsymbol{k}) = \sum_{\boldsymbol{q},ll'} \frac{c'_{mn,ll'}}{\imath\omega-\xi_{\boldsymbol{k}-\boldsymbol{q},l}-\varepsilon_{\boldsymbol{q},l'}} + \frac{d'_{mn,ll'}}{\imath\omega-\xi_{\boldsymbol{k}-\boldsymbol{q},l}+\varepsilon_{-\boldsymbol{q},l'}}.
\end{equation}
The coefficients in the numerators are 
\begin{align}
    c'_{mn,ll'} & = c_{mn,l'} \mathcal{U}_{\boldsymbol{k}-\boldsymbol{q},ml} \mathcal{U}_{\boldsymbol{k}-\boldsymbol{q},ln}^{\dagger} [n_F(-\xi_{\boldsymbol{k}-\boldsymbol{q},l}) + n_B(\varepsilon_{\boldsymbol{q},l'})], \\
    d'_{mn,ll'} & = d_{mn,l'} \mathcal{U}_{\boldsymbol{k}-\boldsymbol{q},ml} \mathcal{U}_{\boldsymbol{k}-\boldsymbol{q},ln}^{\dagger} [n_F(\xi_{\boldsymbol{k}-\boldsymbol{q},l}) + n_B(\varepsilon_{-\boldsymbol{q},l'})],
\end{align}
where $n_F(\xi)=1/(e^{\beta\xi}+1)$ and $n_B(\varepsilon)=1/(e^{\beta\varepsilon}-1)$ are the Fermi and Boson distribution function, respectively. 
Finally, the retarded Green's funtion for electrons is obtaineded via the analytic continuation $\imath\omega\rightarrow\omega+\imath 0^+$
\begin{equation}\label{eq:G_retarded}
    G_{mn}^{R} (\omega,\boldsymbol{k}) = \sum_{\boldsymbol{q},ll'} \frac{c'_{mn,ll'}}{\omega+\imath 0^+-\xi_{\boldsymbol{k}-\boldsymbol{q},l}-\varepsilon_{\boldsymbol{q},l'}} + \frac{d'_{mn,ll'}}{\omega+\imath 0^+-\xi_{\boldsymbol{k}-\boldsymbol{q},l}+\varepsilon_{-\boldsymbol{q},l'}}.
\end{equation}
In the zero-temperature limit, this expression can be approximated by a summation of delta functions 
\begin{equation}
    \sum_{ll'} c'_{mn,ll'}\delta(\omega-\xi_{\boldsymbol{k}-\boldsymbol{q},l}-\varepsilon_{\boldsymbol{q},l'}), \quad \text{for} \quad \xi_{\boldsymbol{k}-\boldsymbol{q},l}<0,
\end{equation}
and 
\begin{equation}
    \sum_{ll'} d'_{mn,ll'}\delta(\omega-\xi_{\boldsymbol{k}-\boldsymbol{q},l}+\varepsilon_{-\boldsymbol{q},l'}) \quad  \text{for} \quad \xi_{\boldsymbol{k}-\boldsymbol{q},l}>0.
\end{equation}

\begin{figure}[t]
    \centering
    \includegraphics[width=0.75\textwidth]{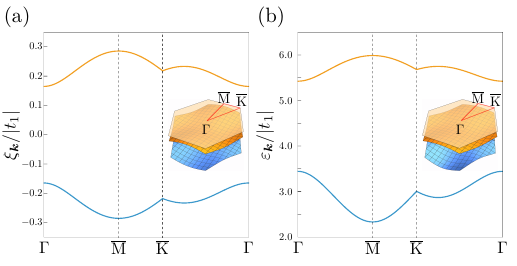}
    \caption{
	Dispersions for (a) spinons and (b) charge bosons along high-symmetry momentum points in the 4-CSL phase.
    Insets shows the full band dispersions in the folded Brillouin zone (top transparent plane) of the $4\boldsymbol{a}_1\times4\boldsymbol{a}_2$ enlarged unit cell.
    The high-symmetry points and lines in momentum space are colored in red.}%
    \label{fig:dispersions}
\end{figure}

\subsubsection*{The electron spectral function of surface CSL}

We now apply the formal expression to compute the ARPES spectrum of the type-II terminated Nb$_3$Br$_8$ based on the hopping parameters derived form our DFT culculations. 
The surface CSLs on the monolayer triangular lattice of Nb$_3$Br$_8$ are stable in the window $2.24 \lesssim U/|t_1| \lesssim 4.146$. 
Within this weak Mott insulator regime, the energy required to excite an electron across the combined spinon and charge rotor gap is accessible via the photoelectric effect. 
Consequently, topological surface states are expected to produce characteristic signatures in ARPES spectra. 
In the ARPES process, the electron spectral function is proportional to the imaginary part of the retarded Green's function after summing over all sublattice indices
\begin{equation}\label{eq:spectra}
    A(\omega \leq 0,\boldsymbol{k}) = - \frac{1}{\pi} \mathrm{Tr}\,\mathrm{Im}\, G^{R}_{mn}(\omega,\boldsymbol{k}).
\end{equation} 

We focus on the $4$-CSL phase where the same CSL is realized in all four sublattice, corresponding to the maximal total Chern number $\pm 4$. 
Becuase of the $\pm \pi/2$ flux on each triangular plaquette formed by the auxiliary field on third nearest neighbor bonds, the unit cell on each sublattice must be doubled along one of the sublattice vectors, as shown in Fig.~3(a) in the main text. 
Nevertheless, the minimal enlarged unit cell required to capture the full $4$-CSL state is of size $4\boldsymbol{a}_1 \times 4\boldsymbol{a}_2$, due to the sublattice structure, which consists of four sites per sublattice. 
The reduced surface Brillouin zone is spanned by the 
\begin{equation}
    \overline{\boldsymbol{b}}_1 = \left(\frac{\pi}{2}, -\frac{\pi}{2\sqrt{3}} \right), \qquad \overline{\boldsymbol{b}}_2 = \left( 0, \frac{\pi}{\sqrt{3}} \right),
\end{equation}
and the high-symmetry point shown in the insets of Fig.~\ref{fig:dispersions} are $\overline{\mathrm{M}} = \overline{\boldsymbol{b}}_2 / 2$ and $\overline{\mathrm{K}} = -\overline{\boldsymbol{b}}_2/3 + \overline{\boldsymbol{b}}_2/3$. 
We adopt this enlarged unit cell to calculate the electron spectral function $A(\omega \leq 0,\boldsymbol{k})$.
Owing to the complete decoupling between sublattices in the low-energy subspace, the calculation for one sublattice is representative of the whole system. 
As shown in Fig~\ref{fig:dispersions}(a), the spinon bands are two-fold degenerate for each spin flavor
\begin{equation}
    \xi_{\boldsymbol{k}}^{\pm} = \pm |\Delta| \zeta_{\boldsymbol{k}} / \sqrt{2},
\end{equation}
where 
\begin{equation}
    \zeta_{\boldsymbol{k}} = \sqrt{3+\cos(4k_x)-2\cos(2k_x)\cos(2\sqrt{3}k_y)}.
\end{equation}
Because of the Hubbard-Stratonovich decoupling, the behavior of charge bosons is closely related to that of spinons. 
Indeed, the two-fold degeneracy also holds for both charge boson bands [see Fig.~\ref{fig:dispersions}(b)]
\begin{align}
    \varepsilon_{\boldsymbol{k}}^{\pm} = \sqrt{2U(\lambda\pm\sqrt{2}|\chi|\zeta_{\boldsymbol{k}})}.
\end{align}

By convolving the Green's functions corresponding to $\xi_{\boldsymbol{k}}^{\pm}$ and $\varepsilon_{\boldsymbol{k}}^{\pm}$, one can obtain the electron spectral function $A(\omega \leq 0,\boldsymbol{k})$. 
Without loss of generality, the Lorentzian broadening parameter is set to $0.01$.
The zero-temperature result, which is approximated by setting $\beta = 100/|t_1|$, for $U/|t_1| = 4.0$ is presented in Fig.~3(e) in the main text. 
Because the charge boson bandwidth is significantly larger than that of spinons [See Fig.~\ref{fig:dispersions}], the spectral weights are almost entirely determined by the second part in Eq.~\eqref{eq:G_retarded}. 
Physically, this means that only the occupied spinons can take part in the ARPES process by exciting a spinon hole at low temperatures. 
Given the energy-conserving condition $\delta(\omega-\xi_{{\boldsymbol{k}-\boldsymbol{q}}}+\varepsilon_{-\boldsymbol{q}})$ and the large energy gap between $\varepsilon_{\boldsymbol{k}}^{\pm}$ compared with the band width of $\xi_{\boldsymbol{k}}^{-}$, the spectral weight is clearly distributed within two energy windows: one around $\omega \approx -3.6 |t_1|$ and the other around $\omega \approx -5.6 |t_1|$, corresponding to transitions involving the lower and upper charge boson bands, respectively. 
In the first (lower charge boson band) window, which is accessible with lower photon energies, the spectral weight consentrates near the $\overline{\mathrm{\Gamma}}$ point and symmetry-related points.
With increasing photon energy, a similar consentration of spectral weight appears near $\overline{\mathrm{K}}$ point and symmetry-related points in the second window. 

\begin{figure}[t]
    \centering
    \includegraphics[width=0.75\textwidth]{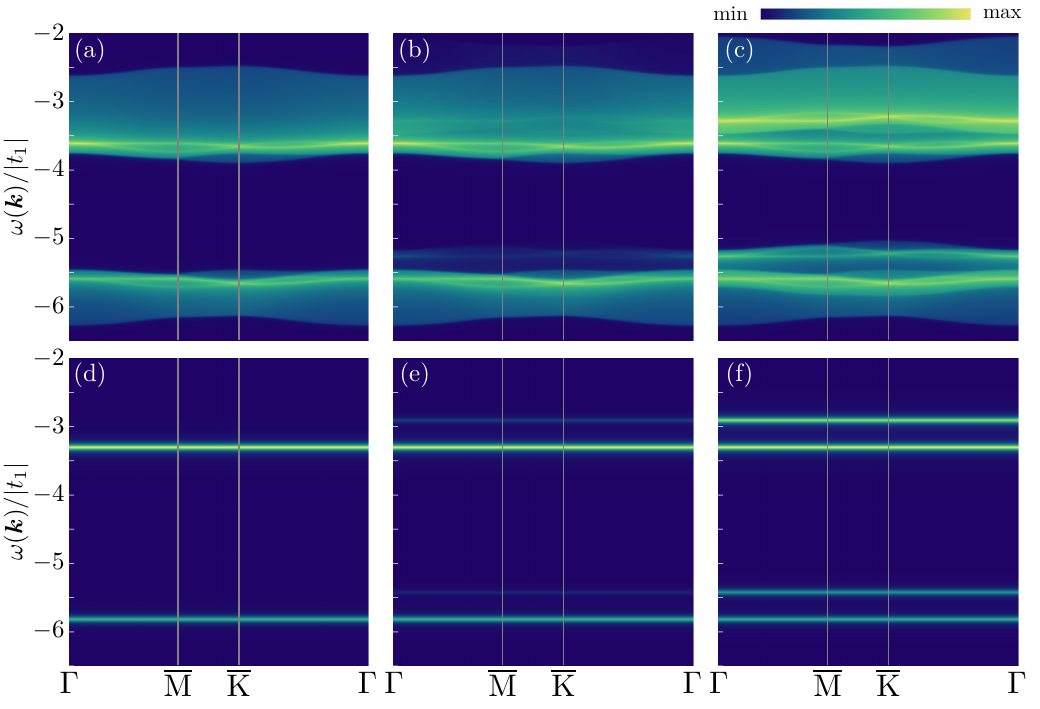}
    \caption{
        The electron spectral functions at $U/|t_1| = 4.0$ for (a-c) 4-CSL and 3-dimer (d-f) phases at different temperatures. 
        From left to right, the temperatures are $\beta/|t_1| = 100$, $\beta/|t_1| = 10$, and $\beta/|t_1| = 1$, respectively. 
        Additional spectral structures appear at about $\omega \approx -5.25|t_1|$ and $\omega \approx -3.25|t_1|$ due to the thermally excited spinons from conduction bands.
    }%
    \label{fig:spectra}
\end{figure}

\begin{figure}[t]
    \centering
    \includegraphics[width=0.75\textwidth]{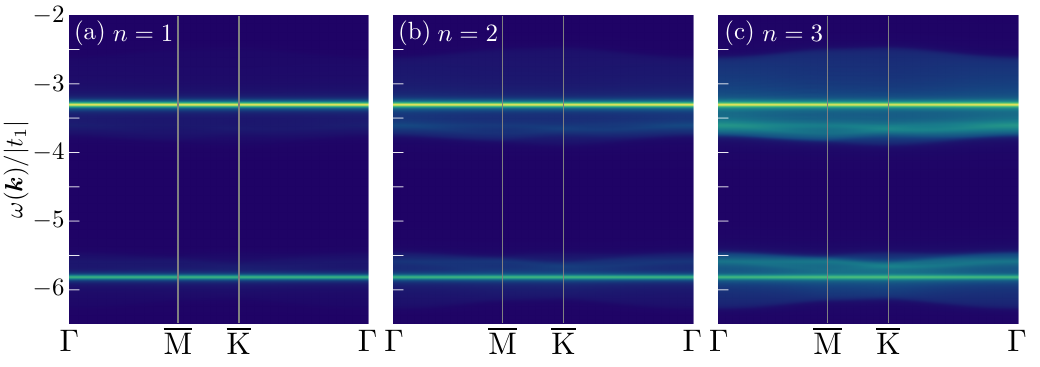}
    \caption{
        The electron spectral functions at $U/|t_1| = 4.0$ for (a) 1-CSL, (b) 2-CSL, and (c) 3-CSL phases at temperatures $\beta/|t_1| = 100$. 
    }%
    \label{fig:spectra-hybrid}
\end{figure}

Figures~\ref{fig:spectra}(a), (b), and (c) show the electron spectral functions $A(\omega \leq 0, \boldsymbol{k})$ in the 4-CSL phase for $\beta/|t_1| = 100$, $\beta/|t_1| = 10$, and $\beta/|t_1| = 1$, respectively.
At finite temperatures, thermal fluctuations promote gapped spinons into the conduction bands $\xi_{\boldsymbol{k}}^{+}$. 
Through the convolution of spinon and charge boson Green's functions, these spinons introduce additional, yet substantially diminished, spectral weights to the electronic spectrum for $\beta/|t_1|=10$ at relatively lower energy $|\omega(\boldsymbol{k})|$. 
These weights are distributed just above the two primary energy windows identified at zero temperature, thereby enriching the spectral structure without altering its fundamental profile. 
The only significant difference occurs at very high temperature $\beta/|t_1|=1$, where the additional spectral weights become more visible. 
Within the temperature regime we considered, the maximum spectral weight always occurs at the point $\overline{\mathrm{\Gamma}}$ (and symmetry-related points) in the energy window near $\omega \approx -3.6|t_1|$.
Consequently, even at elevated temperatures, the characteristic distribution of electron spectral function can still serve as qualitative evidence of the existence of surface CSLs on monolayer Nb$_3$Br$_8$ and 
can be compared with the bilayer counterparts~\cite{Date2025}.
To clearly distinguish the 4-CSL phase from the topologically trivial phase, we present the spectra functions for the 4-dimer phase at the same $U/|t_1|$ and temperatures in Fig.~\ref{fig:spectra}(d-f). 
The dimer configuration on each triangular sublattice is consistent with that in Fig. 3(b) of the main text. 
In this case, the dispersions for both spinons $\xi_{\boldsymbol{k}}^{\pm}$ and charge bosons $\varepsilon_{\boldsymbol{k}}^{\pm}$ become constants. 
Therefore, the convolved spectral functions are two flat bands instead of continua in the zero-temperature limit as shown in Fig.~\ref{fig:spectra}(d). 
As temperature increases, the additional bands with different distributions of spectral weight become pronounced because of the thermally excited spinons on the conduction bands. 
Likewise, the maximum spectral weight also occurs at $\overline{\mathrm{\Gamma}}$ and $\omega \approx - 3.3|t_1|$. 
Finally, to demonstrate the evolution of the electron spectral function as a function of the number of sublattice CSL states, the simulation results for three $n$-CSL phases at the zero-temperature limit are shown in Fig.~\ref{fig:spectra-hybrid}. 
It is obvious that the flat band structure from the $(4-n)$-dimer states persists in all cases, and these contributions to the spectral weight prevail in both energy windows. 
The spectral continua originate from the CSL state, and their spectral weights become increasingly significant as $n$ increases. 
Increasing $n$ to 4 will completely eliminate the contribution from dimers, leaving only spectral continua caused by CSLs, which means the electron spectral function returns to the result shown in Fig.3(e) of the main text.

\bibliography{Ref.bib}